\documentclass[aps,prl,superscriptaddress,floatfix]{revtex4}

\usepackage{bm,bbm,amsmath,amssymb,amsbsy,amsthm,graphicx,subfigure,color,txfonts,multirow}
\usepackage{amsfonts}
\usepackage{array}
 
\usepackage[utf8x]{inputenc}
\usepackage[greek,english]{babel}
\usepackage{teubner}
\newcommand{\ket}[1]{| #1 \rangle}

\newcommand{\ketbra}[2]{| #1 \rangle\langle #2 |}

\newcommand{\integer}[1]{\lfloor #1 \rfloor}

\begin{document}

\title{Weaving and neural complexity in symmetric quantum states}


\author{Cristian E. Susa}

\email{cristiansusa@correo.unicordoba.edu.co}
\affiliation{Departamento de F\'isica y Electr\'onica, Facultad de Ciencias B\'asicas, Universidad de C\'ordoba, Carrera 6 No. 76-103, Monter\'ia, Colombia}

\author{Davide Girolami}

\email{davegirolami@gmail.com}
\affiliation{Los Alamos National Laboratory, Theoretical Division, P.O. Box 1663 Los Alamos, NM, 87545, USA}

\begin{abstract}
We study the behaviour of two different measures of the complexity of multipartite correlation patterns, weaving and neural complexity, for  symmetric quantum states. Weaving is the weighted sum of genuine   multipartite correlations of any order, where the weights are proportional to the correlation order. The neural complexity, originally introduced to characterize correlation patterns  in classical neural networks, is here extended to the quantum scenario. We derive closed formulas of the two quantities for GHZ states mixed with white noise. 

 \end{abstract}
 
 \date{\today}
 \maketitle
 

\section{Introduction}
\label{intro}
Correlations, capturing statistical relations between measurements performed at different times, or at different sites,  take centre stage in many disciplines, as they often unveil dynamical and structural properties of complex systems.  Yet, while bipartite correlations can be assumed to be well understood both in the classical and quantum scenarios, multipartite correlations are still somehow terra incognita, due to the daunting number of degrees of freedom that are necessary to describe systems of many particles \cite{bennett}. We clarify that in this work we mean by ``correlations'' all the statistical dependencies between two or more physical systems. While correlation functions, e.g. covariances, capture linear correlations between variables, we here consider correlation measures to be more general descriptors of the information about joint properties of composite systems.   Correlation patterns, the amount of correlations of different orders (tripartite, four-partite, and so on), describe collective properties of many-body systems, as demonstrated in recent theoretical works \cite{ref1,ref2,ref3,ref4,ref5}, and also verified experimentally \cite{ref6}.  Recently, we proposed a framework to describe {\it genuine} multipartite correlations for quantum and classical systems, providing a method to   unambiguously compute correlations of  order $2\leq k\leq N$ in an $N$-particle system \cite{prl}.  As  states encoding the same amount of information can display very different correlation patterns, we introduced an index to classify them, called weaving. Genuine multipartite correlations express how a many-body system is different from the sum of its parts independently investigated, while weaving captures how such difference scales with the size of the considered parts.
 
In this Special Issue, after recalling their definitions, we provide closed formulas of both genuine multipartite correlations and weaving, as measured by the von Neumann  relative entropy, for the  $N$-qubit GHZ  (Greenberger-Horne-Zeilinger) state  mixed with white noise \cite{GHZs}, a configuration of high relevance for quantum information processing \cite{nielsen}. For such states, we run a comparison between weaving and the quantized neural complexity \cite{tse}, a measure which has been employed to characterize correlation patterns in neural networks.

\section{Genuine multipartite correlations and weaving}
 Genuine multipartite correlations describe emerging joint properties of many-body systems which are intrinsically irreducible to features of the system parts. Specifically, given an $N$-partite (classical or quantum) system, the correlations of order $k$ represent the information which cannot be obtained from clusters of  $k$ or less subsystems. In a recent work, we propose a method to compute genuine multipartite correlations (including classical and quantum contributions) of any order \cite{prl}. An advantage of this approach is that the obtained measure of correlations  is relatively easy to compute. More important,  it meets a set of expected criteria of monotonicity  under local operations.
Let $\rho_N$ be the density matrix representing the state of an $N$-particle quantum system $\mathcal{S}_N$. We define the correlations of order higher than $k, 2\leq k\leq N-1,$ as the information about the total system that is still missing when one has full knowledge of a coarse grained partition $\{\mathcal{S}_{k_1}, \mathcal{S}_{k_2}, \ldots,\mathcal{S}_{k_m}\}, \sum_{i=1}^m k_i=N, k=\max\{k_i\} $, where $\mathcal{S}_{k_i}$ is a cluster of $k_i$ subsystems. From an 
information-theoretic viewpoint, this information can be quantified by the distance of the 
total state to the set of tensor products describing up to $k$-party clusters. Such set reads $\underline{\mathcal{P}}_k=\left\{\sigma_N=\bigotimes_{i=1}^{m}\sigma_{k_i};\;\forall k_i:\,\sum_{i=1}^{m}k_i=N,\, k=\max\{k_i\} \right\},$ where $\sigma_{k_i}$ is the quantum state of a $k_i$-partite cluster. 
These sets form the hierarchy 
$\underline{\mathcal{P}}_1\subset \underline{\mathcal{P}}_{2}\subset\cdots\subset\underline{\mathcal{P}}_{N-1}\subset\underline{\mathcal{P}}_{N},$ where 
$\underline{\mathcal{P}}_{N}$ is the whole Hilbert space of the system. Let us clarify the framework with an example. For $k=1, N=3,$ the set $ \underline{\mathcal{P}}_1=\left\{\sigma_3=\sigma_{[1]}\otimes\sigma_{[2]}\otimes\sigma_{[3]}\right\} $ consists of all the single-particle product states.  For $k=2$, $
\underline{\mathcal{P}}_2=\underline{\mathcal{P}}_1\cup\mathcal{P}_{2},$
where $\mathcal{P}_2$ contains all the products $\{\sigma_2\otimes\sigma_1\}$ obtained by permutations of the subsystems. Hence, $\mathcal{P}_k$ is the set of product states with at least one $k$-partite forming cluster. In particular, $\sigma_2$ is the joint state of two subsystems, e.g., 
$\mathcal{S}_{[1,2]}$ ($\mathcal{S}_{[1]}$ and $\mathcal{S}_{[2]}$). Generally, one has
$\underline{\mathcal{P}}_k=\underline{\mathcal{P}}_{k-1}\cup\mathcal{P}_{k}$,
where $\mathcal{P}_{k}$ is the set of all the possible states with at least one $k$-partite 
term. One can then quantify multipartite correlations higher than $k$ as the geometric distance of the given  state to the set $\underline{\mathcal{P}}_k$. This usually implies a challenging optimization, which is yet significantly simplified by employing the relative entropy $S(\rho||\sigma)=-S(\rho)-\mathrm{Tr}(\rho\log\sigma)$, being 
$S(\rho)=-\mathrm{Tr}(\rho\log\rho)$ the von Neumann entropy. In this case, the closest product state to the global 
state is the tensor product of its marginals \cite{kavan,szalay}. As the von Neumann entropy is subadditive, $S(\rho_i)+S(\rho_j)\geq S(\rho_{ij}), \forall i,j,$  for systems invariant under subsystem permutations, the closest state is always 	$\sigma_N=\left(\bigotimes_{i=1}^{\integer{N/k}}\rho_{k}\right)\otimes\rho_{N\bmod k}$. 
 Hence, the amount of correlations of order higher than $k$ is given by
\begin{eqnarray}	\label{min2}
	S^{k\rightarrow N}(\rho_N):&=&\min_{\sigma_N\in\underline{\mathcal{P}}_k}S(\rho_N||\sigma_{N}) =\sum_{i=1}^{m}S(\rho_{k_i})-S(\rho_N)\\
&=&\integer{N/k}S(\rho_{k})+(1-\delta_{N \bmod k,0})S(\rho_{N\bmod k})-S(\rho_N).\nonumber
\end{eqnarray}
By construction, genuine $k$-partite correlations are then quantified by
\begin{equation}	\label{genuinek}
	S^{k}(\rho_N):=S^{k-1\rightarrow N}(\rho_N)-S^{k\rightarrow N}(\rho_N).
\end{equation}
Note that the total correlations are given by the sum of the correlations of any order, the multi-information $S^{1\rightarrow N}(\rho_N)=\sum_{k=2}^N S^{k}(\rho_N)=\sum_{i=1}^N S(\rho_{[i]})-S(\rho_N)$, a non-negative multipartite generalization of the mutual information. We proved that the measures defined in Eqs.~(\ref{min2}),(\ref{genuinek}), contrary to all the previous proposals (to the best of our knowledge), satisfy a set of expected constraints 
\cite{prl}:
\begin{itemize}\setlength\itemsep{.01em}
	\item Adding a disjoint $n$-partite system,  
cannot create correlations of order higher than $n$, $S^{n\rightarrow N}(\rho_N) \geq S^{n\rightarrow N+n}(\rho_{N+n})$.
	\item Local (single sites) operations, represented by  CPTP (completely positive trace-preserving) maps $\Pi_i\Phi_{[i]}, \Phi_{[i]}=I_1\otimes\ldots \Phi_i\otimes\ldots\otimes I_N,$  cannot create correlations of any order $k$, and cannot increase the amount of correlations higher than any order $k$, $S^{k}(\rho_N)=0\Rightarrow S^{k}(\Pi_i\Phi_{[i]}(\rho_N))=0, \forall k$.
	\item Partial trace of $n$ subsystems cannot increase correlations of order higher than $k<N-n$, $S^{k\rightarrow N}(\rho_N) \geq S^{k\rightarrow N-n}(\rho_{N-n})$.
	\item Distilling $n$ subsystems by fine graining ${\cal S}_{[i]}\rightarrow {\cal S}_{i'}=\{{\cal S}_{[i_j]}\}, j=1,\ldots,n+1$,  cannot create correlations of order higher than $k+n$, for any $k$, $S^{k+n \rightarrow N+n}(\rho_{N+n})=S^{k\rightarrow N}(\rho_N)=0$.
	\item Total correlations are superadditive, $S^{1\rightarrow N}(\rho_N)\geq \sum_{i=1}^m S^{1\rightarrow k_m}(\rho_{k_m})$.
\end{itemize}
While a consistent measure of genuine multipartite correlations is an important tool to investigate many-body systems, computing correlations is insufficient to fully discriminate correlation patterns. In the quantum scenario, it is well known that there exists an infinite amount of {\it kinds} of multipartite entanglement, such that there is not  LOCC (Local Operation and Classical Communication) transformation which can convert a state into another belonging to a different equivalence class (i.e., being entangled in a different way) \cite{cirac}.  Also, the structure of classical networks is not fully captured by measures of correlations \cite{tse,ay1,amari}.
Classifying without amibiguities multipartite systems is a challenging problem. We proposed a potential solution to the issue by introducing weaving, an index assigning different importance to correlations of different order. The idea is to describe how the information missing about the whole system scales when one studies clusters of increasing size.   The relative entropy measure of weaving is given by  
\begin{equation}
    W_S(\rho_N)=\sum_{k=2}^{N}\omega_kS^k(\rho_N)=\sum_{k=1}^{N-1}\Omega_kS^{k\rightarrow N}(\rho_N) ,
    \label{weav}
\end{equation}
where   $\omega_k=\sum_{i=1}^{k-1}\Omega_i, \Omega_k \in \mathbb{R}^+$. The meaning of the weaving measure is 
determined by the choice of weights. For instance, weaving equals total correlations for $\omega_k=1$, 
$\forall k$, while it yields genuine k-partite correlations if $\omega_{l}=\delta_{lk}$, $\forall l$. For any choice of the weights, weaving satisfies, by construction,   the properties of contractivity under local operations, $W_S(\rho_N)\geq W_S(\Pi_i\Phi_{[i]}(\rho_N))$, and additivity, $W_S(\otimes_i \rho_i)=\sum_i W_S(\rho_i)$. We also showed that, by choosing weights proportional to the correlation order, the index is able to rank several classes of correlated classical and quantum states, discriminating among  multipartite states taking the same value of total or N-partite correlations, also being sensitive to the dimension of the subsystems. Hence, weaving appears as  an information-theoretic consistent alternative to the many complexity measures appeared in literature \cite{prl}. For the sake of comparison, we introduced a quantum version of the neural complexity \cite{tse}, proposed for classical  variables,
\begin{eqnarray}\label{comp}
 C(\rho_N):=\sum_{k=1}^{N-1}  k/N C^{(k)}, \qquad C^{(k)}=S^{1\rightarrow N}(\rho_N)-N/k\langle S^{1\rightarrow k}(\rho_k)\rangle,
\end{eqnarray}
 where the average term is computed over the $\binom{N}{k}$  clusters of $k$ subsystems ${\cal S}_k$. Note that each term $C^{(k)}$ measures how much the total correlations on size $k$ clusters $S^{1\rightarrow k}(\rho_k)$ deviate from linearly increasing with the cluster size.   Also, the term  $C^{(1)}$ measures the total correlations in the global state, while  $C^{(N-1)}$ is ($1/(N-1)$ times)  the quantum excess entropy of the state \cite{ay1}. 
The quantity originally aimed at capturing peculiar properties of neural configurations of the visual cortex. While an appealing, computationally friendly measure of the rate of the correlation scaling, the neural complexity, as well as the proposed alternative geometric variants \cite{ay1,amari,jost2,galla,attempt}, does not meet the desirable information-theoretic constraints of contractivity under local manipulation of the system, e.g. it can arbitrarily increase under adding disjointed subsystems \cite{prl}.

\section{Comparative study of weaving and neural complexity}
  
We compare the relative entropy of weaving and the quantum neural complexity for mixtures of white-noise and the  $N$-qubit GHZ   state \cite{GHZs}:
\begin{equation}
	\rho^{\mathrm{GHZ}}_{N}=\frac{p}{2^{N}}I_{2 N}+(1-p)\ketbra{\mathrm{GHZ}_N}{\mathrm{GHZ}_N} , \qquad p\in [0,1],
	\label{GHZN}
\end{equation}
where $I_{2N}$ is the $2N \times 2N$ identity matrix   and 
$\ket{\mathrm{GHZ}_N}=(\ket{0}^{\otimes N}+\ket{1}^{\otimes N})/\sqrt{2}$.
The state is highly symmetric, being invariant under subsystem permutations. Discarding $N-k$ subsystems via partial trace gives $\rho^{\mathrm{GHZ}}_k=\mathrm{Tr}_{\{N-k\}}\rho^{\mathrm{GHZ}}_{N}=\frac{p}{2^k}I_{2k}+\frac{(1-p)}{2}(\ketbra{0}{0}^{\otimes k}+\ketbra{1}{1}^{\otimes k})$. Thus, one has
\begin{eqnarray}
	S^{k\rightarrow N}(\rho^{\mathrm{GHZ}}_N)&=& \integer{N/k}S(\rho^{\mathrm{GHZ}}_{k})+S(\rho^{\mathrm{GHZ}}_{N\bmod k})-S(\rho^{\mathrm{GHZ}}_N),\\
	S(\rho^{\mathrm{GHZ}}_{k})&=& -2\left(\frac{p}{2^{k}}+\frac{1-p}{2}\right)\log{\left(\frac{p}{2^{k}}+\frac{1-p}{2}\right)}-\left(2^{k}-2\right)\frac{p}{2^{k}}\log{\frac{p}{2^{k}}}\nonumber,\\
	 S(\rho^{\mathrm{GHZ}}_{N}) &=& -\left(2^{N}-1\right)\frac{p}{2^{N}}\log{\frac{p}{2^{N}}} 
	-\left(1-\frac{2^{N}-1}{2^{N}}p\right)\log{\left(1-\frac{2^{N}-1}{2^{N}}p\right)},\nonumber\\
	C^{(k)}(\rho^{\mathrm{GHZ}}_{N})&=&\frac{N}{k}S(\rho^{\mathrm{GHZ}}_{k})-S(\rho^{\mathrm{GHZ}}_{N}).\nonumber
\end{eqnarray}

\begin{figure}[h]
 \centering
  \includegraphics[width=16cm,height=24cm]{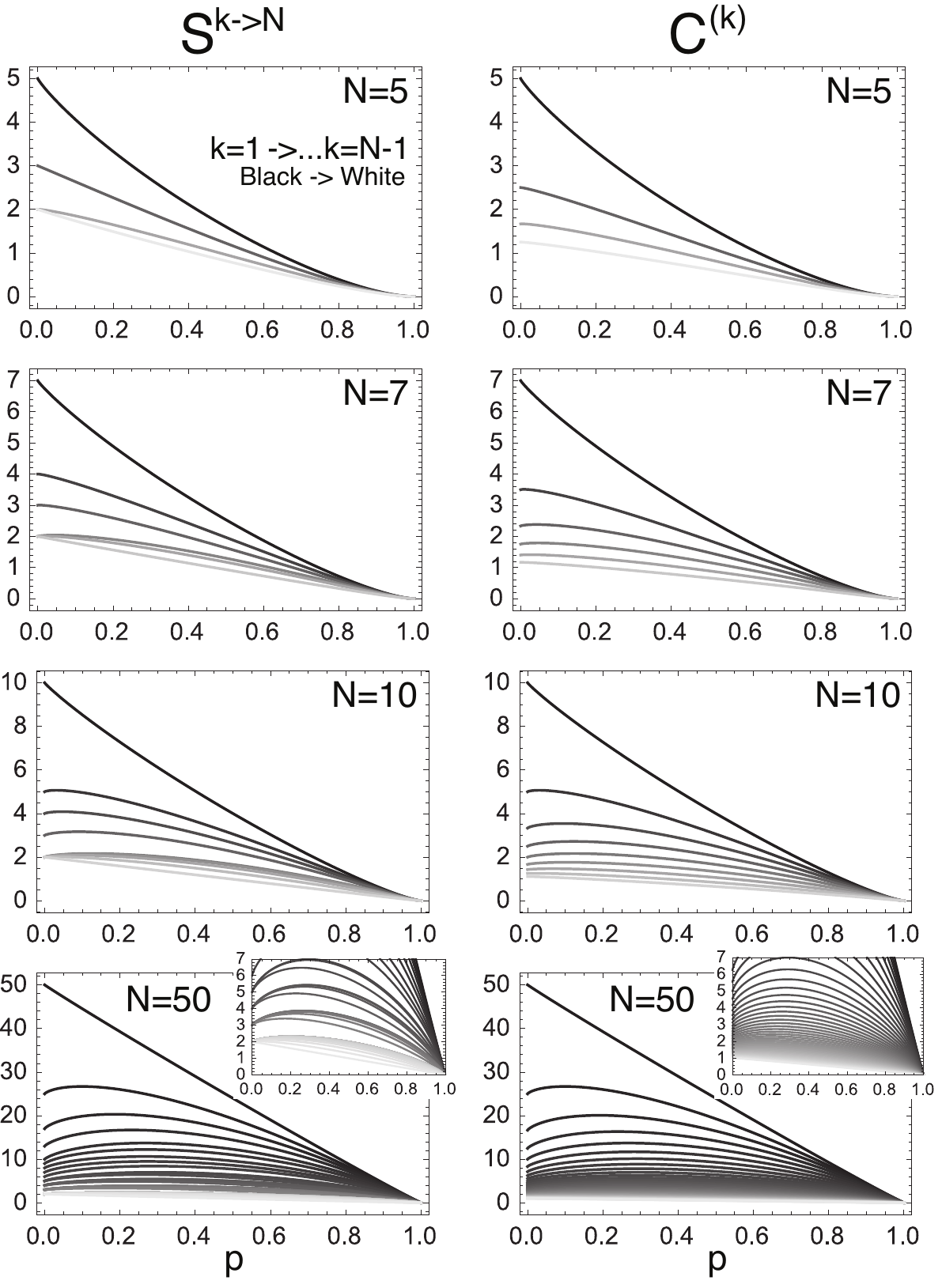}
  \caption{Correlations of order higher than $k$, $S^{k\rightarrow N}$, and components $C^{(k)}$ of the neural complexity, for $N=5,\,7,\,10$ and $50$ qubits. 
  The grey scale goes from black to white as $k$ increases. A zoom of both quantities for the $N=50$ case is shown in the insets.}
  \label{Comfig2}
\end{figure}

\begin{figure}[h]
 \includegraphics[width=8.5cm,height=6cm]{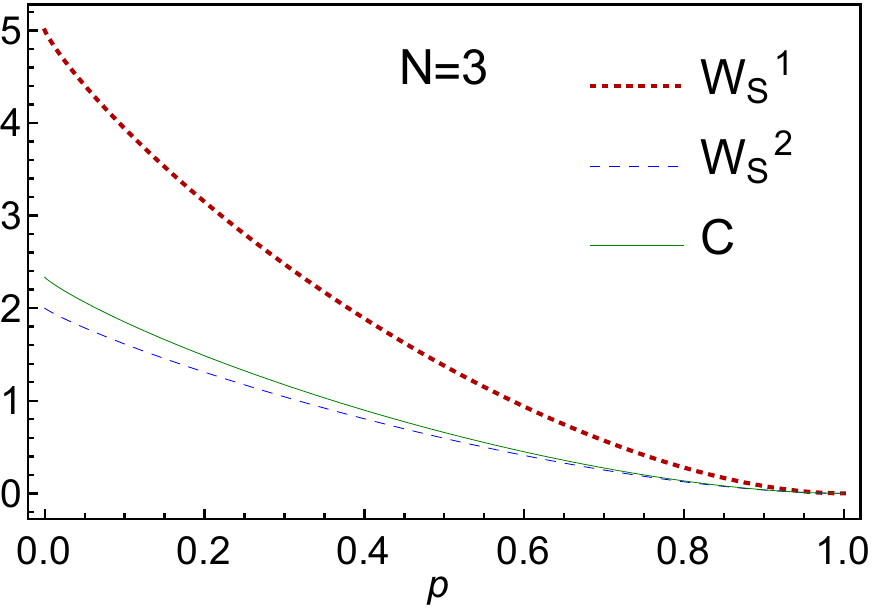}
 \includegraphics[width=8.5cm,height=6cm]{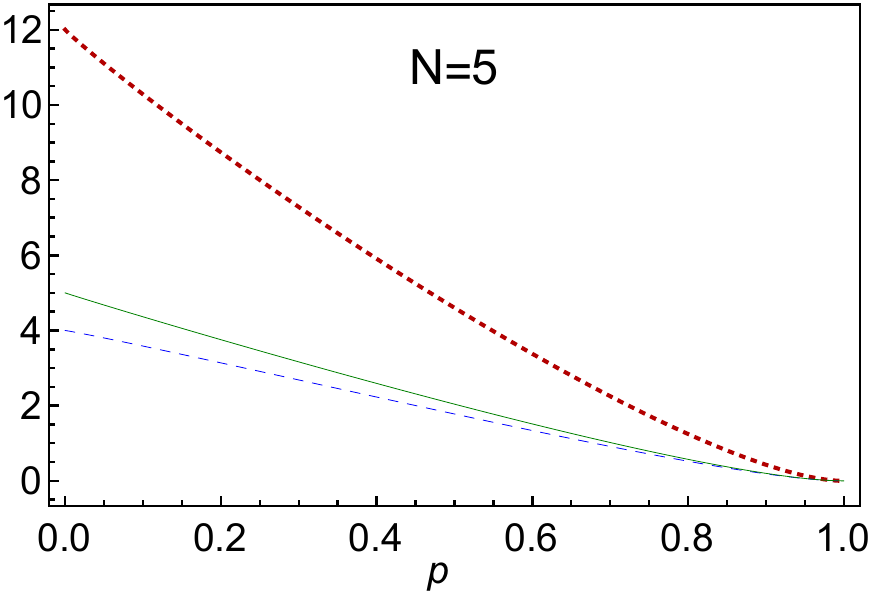}
 \includegraphics[width=8.5cm,height=6cm]{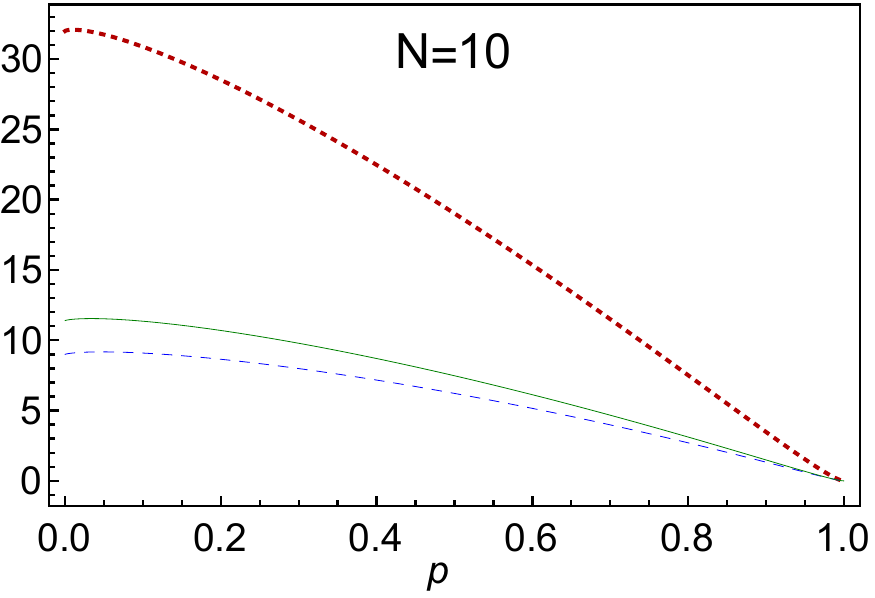}
 \includegraphics[width=8.5cm,height=6cm]{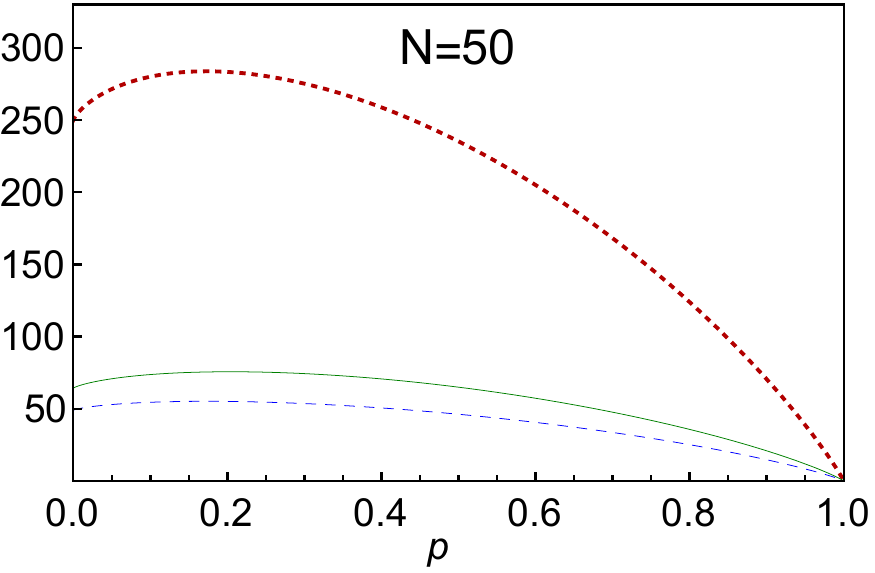}
 \includegraphics[width=8.5cm,height=6cm]{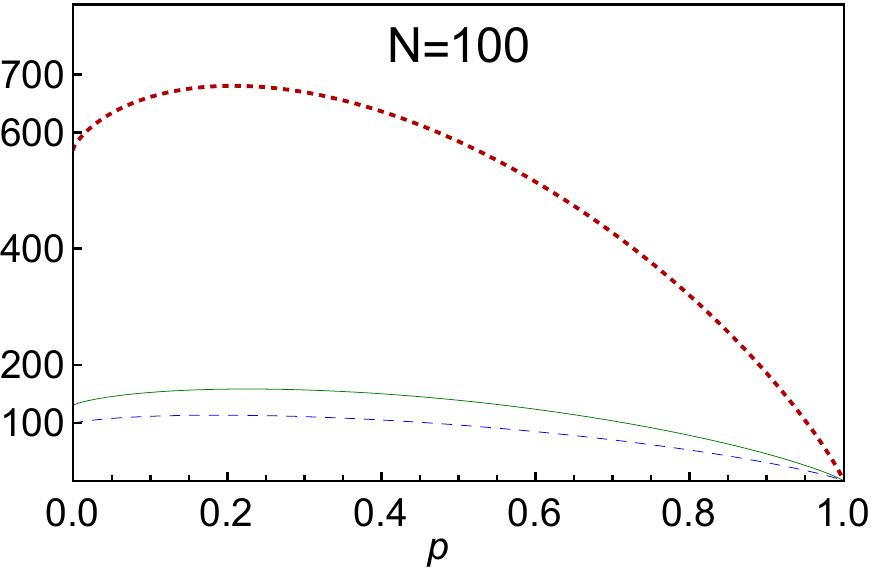}
 \includegraphics[width=8.5cm,height=6cm]{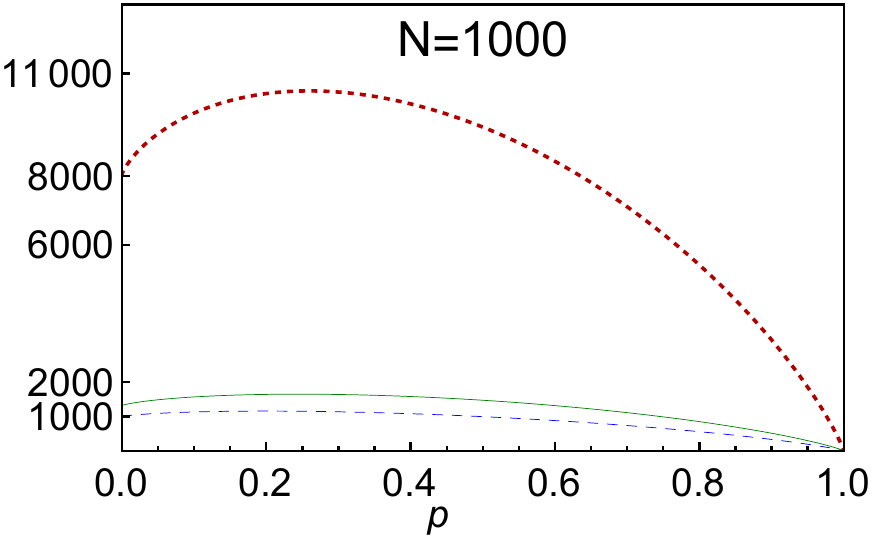}
 \includegraphics[width=8.5cm,height=6cm]{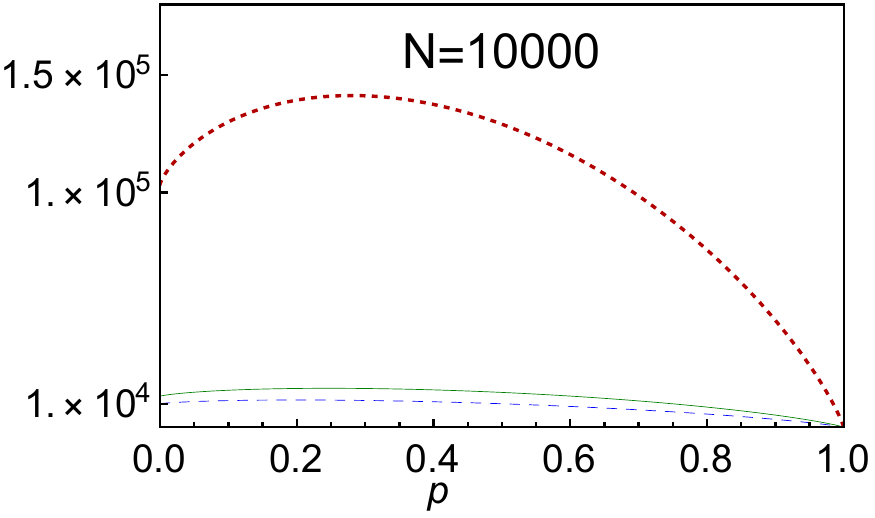}
 \includegraphics[width=8.5cm,height=6cm]{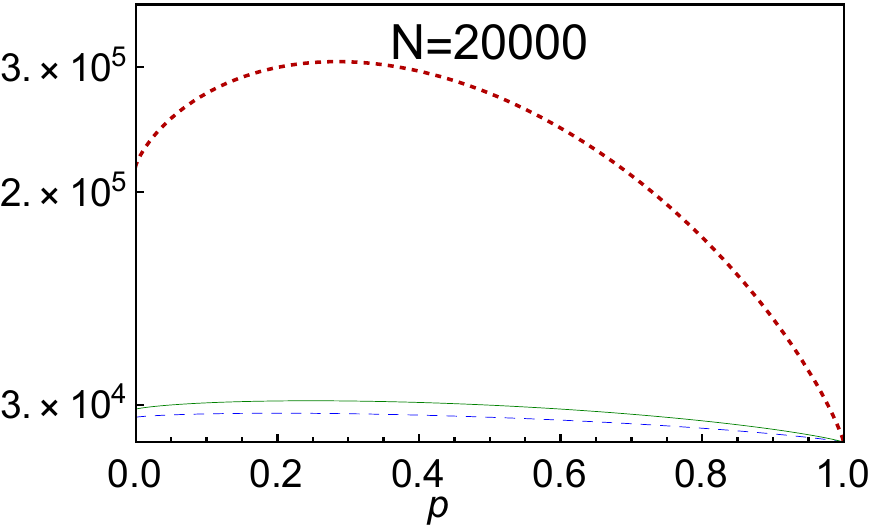}
  \caption{Weaving measures $W^{1,2}_S$, depicted by the red dotted line and the blue dashed line respectively, and neural complexity $C$, green continous line, computed as a function of the mixing parameter $p$, for  depolarized GHZ states of  $N=3,5,10,50,100,1000, 10 000, 20 000$ qubits.}
  \label{Comfig1}
\end{figure}

We perform a numerical comparison between the terms $S^{k\rightarrow N}$ and $C^{(k)}$, by varying the number of particles $N$, reported in Figure \ref{Comfig2}. Although $S^{k\rightarrow N}$ and $C^{(k)}$ have quite similar behaviours, the plot manifests the peculiar correlation structure of the GHZ state, which displays non-zero correlations of order $k$ if only if $\lceil N/(k-1)\rceil\neq \lceil N/k \rceil$, as discussed in Ref.~\cite{prl}, while the components of the neural complexity always take different, non-vanishing values (see the insets for $N=50$).
We then compare  the weaving measures $W^{1,2}_S$ defined in Eq.~(\ref{weav}), for two different choices of the weights, $\Omega_k^{1}=1, \forall k, \Omega_k^{2}=k/N$, respectively, and the neural complexity defined in Eq.~(\ref{comp}).  We stress that complexity measures usually represent a hard computational problem even for states of a few qubits, while weaving and neural complexity are widely applicable because of their expressions are easy to compute.  The results are given in Figure \ref{Comfig1}. By increasing the number of qubits, the weaving measures, as well as the neural complexity, do not reach the maximum in the pure state ($p=0$). This happens as we are here applying a correlated noise and counting both classical and quantum correlations (the state has just $N$-partite quantum correlations), instead of depolarizing independently  each qubit. The same occurs for geometric complexity measures, when computed for Dicke states mixed with white noise \cite{attempt}.  
 

\section{Conclusion}
The compact expressions of complexity measures make them interesting tools to investigate multipartite correlation patterns, which are carriers of information about key properties of  many-body systems. We here reported a comparative study of weaving and neural complexity for correlated symmetric quantum states subject to a white noise channel. We anticipate three lines of research emerging from the interplay between information theory and complexity science concepts. A full-fledged characterization of  genuine multipartite {\it quantum} correlations, including entanglement and other kinds of correlations, after promising results \cite{kavan,giorgi,paula}, may be achieved by adapting the framework here proposed.  Second, our results pave the way for investigating the intriguing notion of genuine multipartite temporal correlations, and therefore the complexity of a process, where the correlations are computed between the state of the system at different times. This promises to be an appealing strategy to characterize memory effects in classical and quantum dynamics. A further line of investigation concerns the experimental detection of quantum complexity without state reconstruction. On this hand, the entropic measure of weaving may be estimated by polynomials of state purities, i.e. trace of squared density matrices and their higher powers. This could be achieved by generalizing  the quantitative bounds to the relative entropy of coherence and coherent information in terms of such functionals  \cite{smolin}.  It is well known that local purities are observables, e.g. via Bell state measurements \cite{ekert,jaksch}.  

\section*{Acknowledgements}
C. E. S. thanks Universidad de C\'ordoba for partial support (CA-097). D. G. is supported at LANL by the LDRD project 20170675PRD2. Part of the work was carried out at the University of Oxford, supported by the EPSRC (Grant
EP/L01405X/1), and the Wolfson College.

\end{document}